\documentclass[twocolumn,preprintnumbers,amsmath,am
ssymb]{revtex4}

\usepackage{graphicx}
\usepackage{dcolumn}
\usepackage{bm}

\begin{document}

\title{ Spectrum of $\pi$ Electrons in Graphene as an Alternant Macromolecule\\
and Its Specific Features in Quantum Conductance }

\author{Alexander Onipko}
\email[]{aleon@ifm.liu.se}
\affiliation{Bogolyubov Institute for Theoretical Physics, 03680 Kyiv, Ukraine}

\date{\today}

\begin{abstract}
An exact description of $\pi$ electrons based on the tight-binding model of graphene as an alternant, plane macromolecule is presented. The model molecule can contain an arbitrary number of benzene rings and has armchair- and zigzag-shaped edges. This suggests an instructive alternative to the most commonly used approach, where the reference is made to the honeycomb lattice periodic in its A and B sublattices. Several advantages of the macromolecule model are demonstrated. The newly derived analytical relations detail our understanding of $\pi$ electron nature in achiral graphene ribbons and carbon tubes and classify these structures as quantum wires. 
\end{abstract}


\maketitle

\section{Introduction}\label{Sec1}
The electronic spectrum of 2D graphite, one-atom thick hypothetical material with the structure of a honeycomb lattice, was first described more, than sixty years ago in the band theory language \cite{Wallace}. Since then it was addressed by many authors, especially after the discovery of multi-wall \cite{I1,I2} and single-wall \cite{I3,CT} carbon tubes and free standing, experimentally accessible graphene \cite{Nov1,Nov2,Nov3,Zhang,Geim}. Two approaches have dominated theoretical modeling of graphene and its daughter structures \cite{Geim}. One following the line of the Wallace pioneer work produced a considerable development based on the Solid State theory methods \cite{Saito-2D,Nakada,Mint3}, see also \cite{Saito,Mint1} and references therein. The other approach has focused on peculiar electronic properties near the Fermi energy, where the methodology of quantum electrodynamics proved to be both heuristic and instrumental \cite{Gus2}. 

The molecular approach to the description of graphene has received comparatively little attention. Zigzag carbon tubes can be thought as build up of cyclacenes \cite{Houk}; similarly, a graphene sheet can be constructed from linear acenes. From this point of view, graphene is a typical alternant macromolecule \cite{P,Low,GR-Fam} representing a vast  field which has been contributed in a number of fundamental studies \cite{P,Low,LJ,LH,SSH,Kiv}.

Recently, simple tight-binding model of $\pi$ electrons in graphene as a macromolecule was briefly reported \cite{Lyuba1}. It was demonstrated that such an approach provides a deep insight into the graphene electronic structure. Several developments have been outlined which were inaccessible or missed in the previous studies. In particular, a controversy between the obvious nonequivalence  of armchair and zigzag directions in the honeycomb lattice and the identity of band structures near the Fermi energy for armchair and zigzag carbon tubes is resolved by exposing principal distinctions between different types of metallic graphene ribbons and related carbon tubes in accurate analytical expressions. 

The purpose of this article is to give a detailed description of the model proposed in \cite{Lyuba1} and partly, its further developments \cite{Lyuba2,Lyuba3,Lyuba4,Lyuba5}. In Section II supplemented by Appendixes A and B, the solution of the Schr\"odinger problem for the graphene sheet framed by armchair and zigzag edges is obtained. It comes to the dispersion relation, where wave vector components are not independent but interrelated via a transcendent equation, typical for alternant oligomers. The usage of this pair of equations is specified for graphene daughter structures, armchair and zigzag carbon tubes and parent graphene ribbons, as the basis for further analysis. The Fermi energy region receives much of attention in Sec. III and Appendix C, where $\pi$ electron spectra of achiral graphene ribbons and carbon tubes are expressed in terms of elementary functions. As a particular application of these results, quantum conductance of graphene-based wires is discussed in Sec. IV. This is followed by a concluding Section. 

\section{Exact Solution of the Schr\"odinger problem}

Shown in Fig.~1 is a honeycomb $N$$\times$$\cal N$ lattice. The lattice label indicates that in the armchair direction, the sheet of graphene contains $N$ hexagons in polyparaphenylene-like chains, whereas in the zigzag direction, it has ${\cal N}$ hexagons forming acene chains. Hydrogen atoms along edges are not shown and not taken into account in the nearest-neighbor tight-binding Hamiltonian $H$. By exploiting $m,n$, and $\alpha=l,\lambda,\rho,r$ labeling explained in Fig.~1, the $\pi$ electron wave function that satisfies the Schr\"odinger equation, 
\begin{equation}\label{1}
H\Psi=E\Psi, 
\end{equation}
can be represented in the form of expansion
\begin{equation}\label{2}
\Psi = \sum_{m}\sum_{n}\sum_{\alpha=l,r,\lambda,\rho}
\psi_{m,n,\alpha} |m,n,\alpha\rangle,
\end{equation}
where $|m,n,\alpha\rangle$ is the $2p_z$ orbital at the $\alpha$th atom of benzene ring with coordinates $\{m,n\}$, the summation is running over all sites of the honeycomb lattice, 
\begin{equation}\label{3}
\psi_{m,n,\alpha}  = \left\{
\begin{array}{ll}
\sum_{j=1}^{\cal N} \phi_{n,\alpha}^j 
\sin \frac{\pi j m}{{\cal N}+1} &\alpha =l,r,\\\\ 
\sum_{j=1}^{{\cal N}+1} \phi_{n,\alpha}^j 
\sin\frac{\pi j (m-1/2)}{{\cal N}+1}&\alpha = \lambda, \rho,
\end{array}\right.
\end{equation}
and unknowns $\phi_{n,\alpha}^j$, $j=1,2,...,{\cal N}$, are to be found from Eq. (\ref{1}). 

Details are described in Appendixes A and B, where solutions to Eq. (\ref{1}) are obtained for the open boundaries, that is for the lattice terminated by armchair and zigzag edges as they appear in Fig.~1. For periodic boundary conditions in $x$ direction, the wave function is also found. 

In Appendix A, the $\pi$ electron spectrum is shown to be determined by equation (energy is in units of the hopping integral $|t|$ \cite{Wallace,Saito})
\begin{equation}\label{4}
E_j^{\pm\;2}=
1\pm4\cos \frac{\pi j}{2({\cal N}+1)}  \cos \frac{\kappa_j^\pm}{2}+
4\cos^2 \frac{\pi j}{2({\cal N}+1)} , 
\end{equation}
where for the sign + and $-$, and each value of $j$, $N$ values of $\kappa_j^\pm$, $ \kappa^\pm_{j,\nu_j}$, $\nu_j=0,1,...,N$-1, are solutions to equation
\begin{equation}\label{5}
\frac{\sin \kappa_j^\pm N}{\sin \kappa_j^\pm (N+1/2)}=
\mp2\cos\frac{\pi j}{2({\cal N}+1)}.
\end{equation}

\begin{figure}
\includegraphics[width=0.35\textwidth]{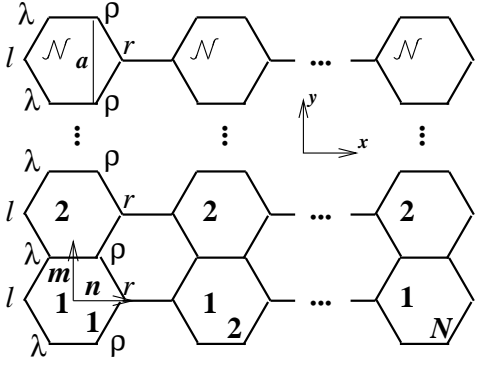}
\caption{Indication of labels of carbon atoms used in the present description of the $\pi$ electron spectrum of $N\times\cal N$  honeycomb lattice. $a$ is the minimal translation distance in the lattice. }
\label{Fig1}
\end{figure}

Thus, $\pi$ electron states of graphene can be classified into 2$\cal N$ "$j$-minus" and "$j$-plus" conduction and equal number of valence bands with $N$ levels $E_j^\pm(\kappa^\pm_{j,\nu_j})$ within each band. This is $4N\cal N$ of the total number $2N(2{\cal N}+1)$ of $\pi$ electron levels. Additionally, there are two $N$-fold degenerate levels with energies $\pm1$. These originate from the states with zero wave-function amplitudes at the $l$ and $r$ sites in linear acenes \cite{Lyuba4}. 

Equation (\ref{5}) makes one quantum number dependent on the other. It appears because of zigzag-shaped edges. Before finding solutions to this equation, let us consider related structures, where zigzag edges either do not exist or their effect can be disregarded. For such daughter structures of graphene, the spectrum is completely determined by an appropriately modified Eq. (\ref{4}) or its substitute Eq.~(\ref{A14}) in the case of periodic boundary conditions in $y$ direction. 

In a graphene strip with $N >>\cal N$, zigzag edges affect a small part of the spectrum. In the limit $N\rightarrow\infty$, these edges play no role at all and $\kappa_j^\pm$ can be replaced by a continuous variable, the wave vector in units of the inverse periodicity $\sqrt{3}a$ in armchair direction, see Fig. 1. 

Hence, the spectrum of an infinitely long graphene strip with armchair edges, called henceforth armchair graphene ribbon (aGR), is fully determined by a single equation that reads
\begin{equation}\label{6}
\begin{array}{l}
E_j^{^{\rm aGR}}(k^\pm_x) =\\
0\le\sqrt{3}k^\pm_x\le\pi\\
\pm \sqrt{
1\pm4\cos \frac{\pi j}{2({\cal N}+1)}  \cos \frac{\sqrt{3}k^\pm_x}{2}+
4\cos^2\frac{\pi j}{2({\cal N}+1)} }\,.
\end{array} 
\end{equation}
Here and in what follows, we are using dimensionless wave vectors in units of $a^{-1}$.

As shown in Appendix A, a pair of equations similar to Eqs. (\ref{4}) and (\ref{5}) can be obtained for periodic boundary conditions in $y$ direction. From Eq.~(\ref{A14}), it follows that the spectrum of zigzag carbon tube with an infinite length [or, simply, zigzag carbon tube (zCT)] is determined by 
\begin{equation}\label{7}
\begin{array}{l}
E_j^{^{\rm zCT}}(k^\pm_x) =\pm \sqrt{
1\pm4\left|\cos \frac{\pi j}{\cal N}\right| \cos \frac{\sqrt{3}k^\pm_x}{2} +
4\cos^2\frac{\pi j}{\cal N} }\,.\\
j=0,1,...,{\cal N}{\rm-1}
\end{array}
\end{equation}

\begin{figure*}
\includegraphics[width=0.9\textwidth,height=0.5\linewidth]{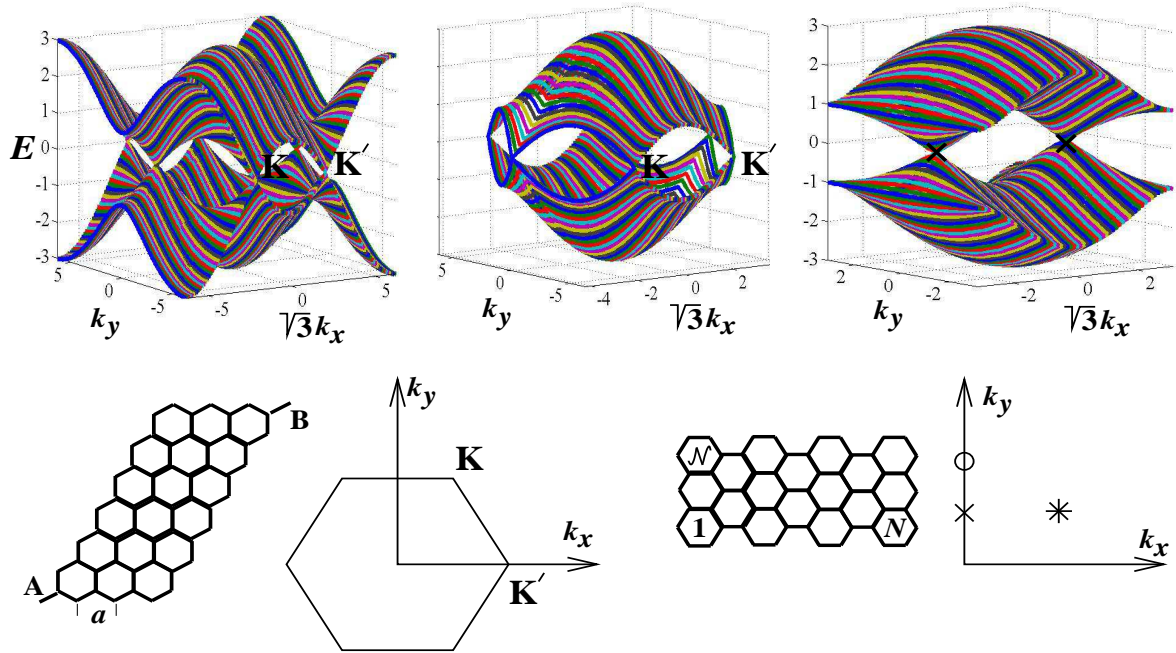} 
\caption{Lower part: 2D graphite lattice build up of A and B triangular lattices (on the left) and $N$$\times$$\cal N$ honeycomb lattice of graphene sheet (on the right). Zero-energy points, six for 2D graphite and two for graphene, are indicated as K and cross points, respectively. Circle $(0,\frac{4\pi}{3})$ and star $(\frac{2\pi}{\sqrt{3}},\frac{2\pi}{3})$ indicate additional zero-energy points of zCT dispersion, Eq. (\ref{7}),  and aCT dispersion, Eq. (\ref{9}), respectively. Distance in k space between cross points (= $\frac{4\pi}{3}$) differs from that is between K and K$'$ points which is equal to $\frac{4\pi}{\sqrt{3}}$. Upper part: dispersion in graphene [on the right, Eq. (\ref{10})] and in 2D graphite [on the left, Eq. (\ref{11})]; mid panel shows the Brillouin zone in 2D graphite. Energy is in units of $|t|$, wave vector is in units of $a^{-1}$. }
\label{Fig2}
\end{figure*}

In the next case, complications of finding the spectrum connected with the necessity to solve Eq. (\ref{5}) are avoided due to that the boundary conditions at zigzag edges are chosen to be periodic. In combination with armchair open edges, this corresponds to a segment of armchair carbon tube (aCTS), the spectrum of which can be written as
\begin{equation}\label{8}
\begin{array}{l}
E_{jj'}^\pm\equiv E_{jj'}^{^{\rm aCTS}} =\\
\pm\sqrt{
1\pm4\cos \frac{\pi j}{2({\cal N}+1)} \left| \cos \frac{\pi j'}{N}\right|+
4\cos^2\frac{\pi j}{2({\cal N}+1)}}\,.\\ 
j=1,2,...,{\cal N},\,j'=0,1,...,N{\rm-1} 
\end{array}
\end{equation}

In the limit ${\cal N}\rightarrow \infty$, one can obtain the band spectrum of armchair carbon tube (aCT) by making a replacement $\frac{\pi j}{{\cal N}+1} \Rightarrow k^\pm_y$ in the above equation. This yields 
\begin{equation}\label{9}
\begin{array}{l}
E_{j'}^{^{\rm aCT}}(k^\pm_y) =\pm
\sqrt{1\pm4\cos \frac{k^\pm_y}{2} \left| \cos \frac{\pi j'}{N}\right|+4\cos^2\frac{k^\pm_y}{2}}.\\
0\le k^\pm_y\le\pi
\end{array}
\end{equation}

Finally, in the limit {$N,{\cal N}\rightarrow\infty$}, when both quantum numbers in Eq. (\ref{4}) can be treated as continuous variables, the dispersion relation for $\pi$ electrons in the infinite graphene sheet reads 
\begin{equation}\label{10}
\begin{array}{l}
E(k^\pm_x,k^\pm_y) =\pm \sqrt{
1\pm4\cos \frac{k^\pm_y}{2}  \cos \frac{\sqrt{3}k^\pm_x}{2}+4\cos^2\frac{k^\pm_y}{2} }.\\
0\le \sqrt{3}k^\pm_x,k^\pm_y\le\pi
\end{array} 
\end{equation}

Usually, another form of dispersion relation is referred identically. Obtained for the honeycomb lattice consisting of two periodic triangular lattices A and B \cite{Wallace}, it reads \cite{Saito,Mint1,Thomsen,Ando,Review}
\begin{equation}\label{11}
\begin{array}{l}
E_{\rm W}(k_x,k_y) =\pm\sqrt{
1+4\cos\frac{k_x}{2}\cos\frac{\sqrt{3}k_y}{2}+4\cos^2\frac{k_x}{2}}\,.\\
0\le k_x,\sqrt{3}k_y\le2\pi
\end{array} 
\end{equation}
The dispersion calculated according these two energy-wave vector relations, one for $N$$\times$$\cal N$ and the other for periodic honeycomb lattice, is shown in Fig. 2. The striking difference between the dispersion in the left and right panels originates from the different symmetry of locally identical lattices.

The subsequent discussion is primarily focused on the energy region near the Fermi energy (for this model, it equals zero). In the majority of related studies, this region associates with zero-energy points of the Wallace dispersion relation (\ref{11}). Therefore, it is worthwhile to compare such points in Eqs. (\ref{10}) and (\ref{11}). These are shown in Fig. 2 and indicated by crosses and by K, K$'$, respectively. K points correspond to the corners of the first Brillouin zone of 2D graphite (middle panel in Fig.~2). Cross points have different coordinates and the number of these zero-energy points is only two. The distance between cross points in the k space differs from that is between neighboring K and K$'$ points. All this is a consequence of the lower symmetry of $N$$\times$$\cal N$ honeycomb lattice in comparison with its counterpart, 2D graphite lattice periodic in A and B sublattices. 

For carbon tubes, zero-energy points of dispersion in zCT, Eq. (\ref{7}), and in aCT, Eq. (\ref{9}), are, respectively, cross/circle points and cross/star points. The dispersion of aGR (\ref{6}) has only one zero-energy point, it is the cross point, and for the zGR dispersion, this point cannot be shown in the real $k_x$-$k_y$ plane. Different manifestations of zero-energy points, as summarized in Table 1, in the band spectra of achiral carbon tubes and graphene ribbons will be illustrated below by concrete examples.

\begin{table}[h]
\caption{ Coordinates of zero-energy points (z.e.p.) in k plane for achiral carbon tubes and graphene ribbons. (a) for zigzag graphene ribbons, zero-energy point cannot be defined in real coordinates $k_x$ and $k_y$; values of the corresponding longitudinal wave vector (in units of $a^{-1}$) at zero-energy point (or points) are shown in column $k_0$. $g^\sigma_\nu$ is degeneracy of the $\nu$th band energy $|E^\sigma_\nu(k_\nu^\sigma=0)|$; (b) $({\cal N}+1)/3$ is an integer; (c), (d) ${\cal N}$ is even, odd. $E^\sigma_{_{\rm H{\rm-}L}}$ denotes the HOMO-LUMO gap in carbon tube segments (CTS) and graphene ribbon strips (GS), see text. } 
\begin{tabular}{cccccc} 
\hline 
$\sigma$ &z.e.p. &$k_x,k_y$  &$k_0$&$g^\sigma_\nu$&$E^\sigma_{_{{\rm H}{\rm-L}}}$\\ 
& $\times$ & 0,$\frac{2\pi}{3}$  & &&  \\
aCT &    & &$\frac{2\pi}{3}$ &  
$\begin{smallmatrix} 1(\nu=0) \\
2(\nu\ne0) \end{smallmatrix}$
&   $\begin{smallmatrix} 0^{\rm (b)} \\
\frac{\pi}{\sqrt{3}({\cal N}+1)}  \end{smallmatrix}$
 \\
& * &$\frac{2\pi}{\sqrt{3}},\frac{2\pi}{3}$&  & &
\\\\
&$\times$ & 0,$\frac{2\pi}{3}$  & & &       \\
zCT&&&0& $2$ & 
$\begin{smallmatrix} 0^{(\rm c)} \\
2{e^{-2N\left|\ln\frac{\pi}{\cal N}\right|}}^{(\rm d)} \end{smallmatrix} $
\\
&o&0,$\frac{4\pi}{3}$& &  &\\\\
aGR&$\times$& 0,$\frac{2\pi}{3}$ & 0 &1 &$2e^{-2N\left|\ln\frac{\pi}{{\cal N}{\rm+1}}\right|}$  \\\\
zGR& & \small{(a)}&$\pi$&1& same \\ 
\hline
\end{tabular}
\label{Table1}
\end{table}

\section{Band Structure Near the Fermi Energy}

In this section, we take a closer look at the spectra of graphene daughter structures specified above, namely, finite-width graphene strips which are infinite in armchair (aGR) and zigzag (zGR) directions, and corresponding carbon tubes which can be viewed as zGR ruled in armchair direction (aCT) and aGR ruled in zigzag direction (zCT). Note that in this description, periodic boundary conditions along the graphene ribbon and respective carbon tube are not used. Such boundary conditions are relevant to a GR ring and CT toroid, the structures which have not the same spectra as the counterparts with open ends \cite{Zhu}.

For carbon tubes, approximate but sufficiently accurate expressions will be derived from the exact dispersion relations. It will be demonstrated that in comparison with previously suggested approximations, these  expressions improve agreement with the exact model results and expose essential differences between the spectra of aCT and metallic zCT. For graphene ribbons, the obtained results extend and correct earlier descriptions.
Also, it seems advantageous that the spectra of finite and infinite graphene ribbons and carbon tubes are described here on equal footings and with the use of the same basic equations. In the case of carbon tubes, our approach leads to the same spectra as were obtained in the framework of zone folding technique \cite{Saito,Saito-2D}. However, the quoted technique is not applicable to graphene ribbons that has motivated the appearance of other calculation schemes to be discussed later on. In comparison with our previous related publication \cite{Lyuba3}, this section places a special emphasis on the origins and values of band degeneracy in graphene ribbons and carbon tubes.

Below, only a part of the spectrum is in focus. It corresponds to the minus branch of Eq. (\ref{4}). 
Therefore, we omit indication "$-$" in energy and wave vector notations. Index $\sigma$ will be used for system labeling, $\sigma\equiv$ aCT, zCT, and so on. 

\subsection{Spectrum of Armchair Carbon Tube}

\begin{figure}
\centering
\includegraphics[width=0.4\textwidth,height=0.7\linewidth]{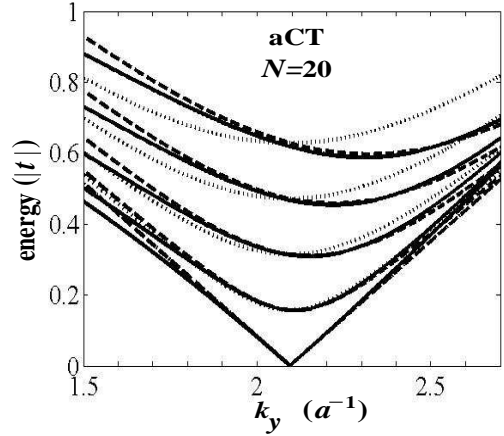}
\caption{Band structure of armchair carbon tubes as calculated from exact equation (\ref{9}), solid lines, and its approximations (\ref{12}), dashed lines, and (\ref{12}) without the $q$ term in parentheses, dotted lines; $N$ = 20. Only few conduction bands are shown. The valence bands with the same band index are just a mirror reflection of conduction bands to negative energies. Spectrum refers to the cross and star points; star point itself does not belong to the spectrum, see text.}
\label{Fig4}
\end{figure}

\begin{figure}
\includegraphics[width=0.5\textwidth,height=0.95\linewidth]{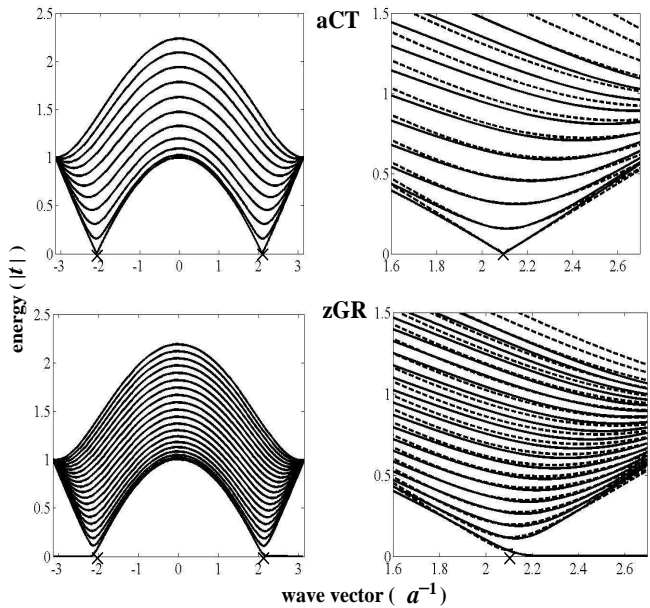}
\caption{Full spectrum (on the left) and its view near cross point (on the right) for armchair carbon tube (aCT) and zigzag graphene ribbon (zGR). $N=20$; solid lines are calculated according Eq. (\ref{9}); dashed lines represent approximation (\ref{12}) for aCT and approximation (\ref{28}) for zGR. Only conduction bands are shown.}
\label{Fig5}
\end{figure}

To obtain an expression of aCT energy spectrum near the Fermi energy, Eq. (\ref{9}) needs to be expanded in power series near the cross and star points. Up to the lowest order in $q=k_y-\frac{2\pi}{3}$, $\frac{\pi j'}{N}$, and $q$, $\frac{\pi (N-j')}{N}$, the minus branch of Eq. (\ref{9}) takes the form
\begin{equation} \label{12}
\begin{array}{ll}
E^\sigma_\nu(q)= {\displaystyle \pm \frac{\sqrt{3}}{2}}
\sqrt{ m^{\sigma\,2}_\nu\left(1-\frac{\sqrt{3}}{2}q\right) + q^2 } ,&\\
&\sigma={\rm aCT}, \\
\displaystyle  m^\sigma_\nu= \frac{2\pi |\nu|}{\sqrt{3}N}<<1, \quad |q|<<1,\,
\nu\rm=0,\pm1,...\,.&
\end{array} 
\end{equation}
Note that Eq. (\ref{12}) represents expansions near both zero-energy points. The state with $\nu,q$ = 0 is not degenerate because the star point does not belong to the aCT spectrum. All other band energies in this spectrum are twofold degenerate. Electron states with energies other, than the band bottoms or tops, are two- and fourfold degenerate for $\nu$=0 and $\nu$$\ne$0 bands, respectively.

Conduction (valence) band bottoms (tops) are attained at 
\begin{equation} \label{12a}
q= q_\nu^{\rm aCT}=\frac{\sqrt{3}}{4} m^{\rm aCT\;2}_\nu, 
\end{equation}
and are equal to
\begin{equation} \label{13}
\left|E_\nu^{ ^{\rm aCT} }(q_\nu^{\rm aCT})\right|= 
\frac{\sqrt{3}}{2}m^{\rm aCT}_\nu \left(1-\frac{3}{32}m_\nu^{\rm aCT\;2}\right).
\end{equation}
The quantity $|\nu|^{-1}\frac{\sqrt{3}}{2}m^{\rm aCT}_\nu$ would be the band spacing, if linear in $q$ term under the root in Eq. (\ref{12}) were disregarded, as, e.g., in Refs. \cite{Ando,Mint1,Mint,Thomsen}. 

In the spirit of analogy to be discussed in Sec. 3E, we refer to the approximation $E^\sigma_\nu(q)= \pm \frac{\sqrt{3}}{2}\sqrt{ m^{\sigma\,2}_\nu + q^2 }$ as relativistic-like approximation. The band structure of aCT calculated with the use of exact Eq. (\ref{9}) and approximately, according to Eq. (\ref{12}), and in the relativistic-like approximation just mentioned, is represented in Fig. 3. Equation (\ref{12})  provides a very accurate reproduction of exact results which need no comments. In contrast, the spectrum in the relativistic-like approximation shown by dotted lines is noticeably worse. However, both approximations are practically equivalent for the description of quantum conductance, and integral characteristics such as density of states.

According to its derivation, spectrum (\ref{12}) refers to the cross and star points. However, the star point does not belong to the spectrum. Thus, with an account to spin degeneracy, the electron state with energy $E^{^{\rm aCT}}_0(0)$ is twofold degenerate, and states with energies $E^{^{\rm aCT}}_{\nu\rm\ne0}(0)$ are fourfold degenerate. 

The same conclusion follows from zone folding \cite{Saito,Mint1,Review,Thomsen,Ando}. However, the band spectrum $E^{^{\rm aCT}}_\nu(k_y)$ shown in Fig. 4 associates with K and K$'$ valleys, and not with the cross and circle points. In our opinion, this is misleading. The distance between these points is different and, as discussed above, K points are characteristic for 2D graphite and not for carbon tubes and graphene ribbons.

\subsection{Spectrum of Zigzag Carbon Tube}

\begin{figure*}
\centering
\includegraphics[width=1.0\textwidth]{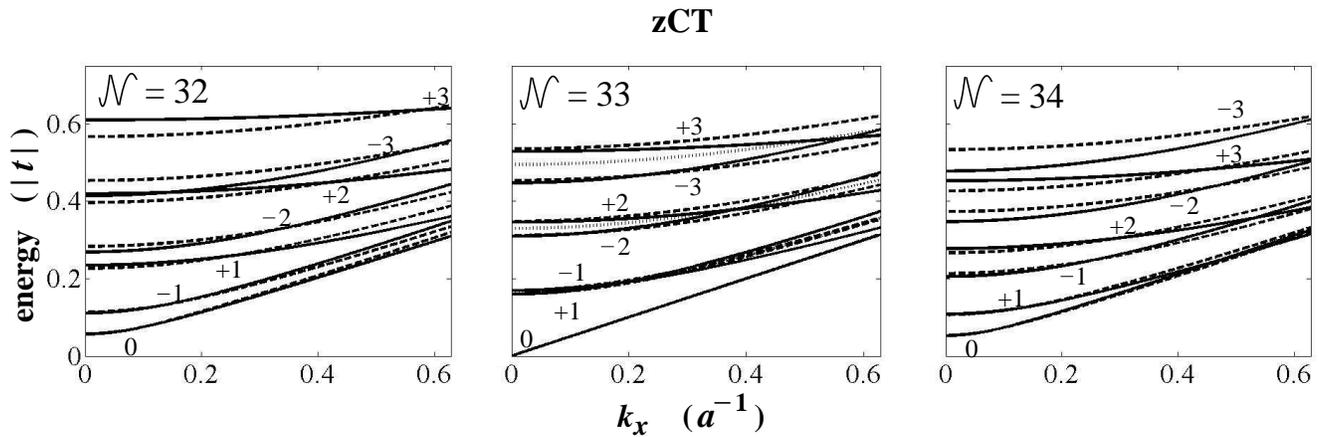}
\caption{Same calculations as in Fig. 3, but for zigzag carbon tubes, semiconducting ${\cal N}=32,34$ and metallic ${\cal N}=33$. Solid and dashed curves labeled by $\nu$ values represent Eqs. (\ref{7}) and (\ref{14}), respectively. Dotted curves in the mid panel correspond to $m_\nu^{\rm zCTm}=\frac{2\pi|\nu|}{{\cal N}}$. Spectra refer to the cross and circle points in Fig. 2.}
\label{Fig6}
\end{figure*}

An analytical expression for zCT spectrum in the vicinity of cross and circle points can be obtained from Eq.~(\ref{7}) in a way similar to the above consideration. We note first that as known, distinct from armchair carbon tubes which are always metallic, zCT has a gapless spectrum, if $j^*={\cal N}/3$ is an integer, $E^{^{\rm zCT}}_{j=j^*}(k_x$=0) = 0. Otherwise, zCT spectrum has a gap. If ${\cal N}/3$ is not an integer, band index of the lowest conduction (highest valence) band can be equal either to $j^*$=$\frac{{\cal N}-1}{3}$ or to $j^*$=$\frac{{\cal N}+1}{3}$. In general, these two possibilities correspond to different types of semiconducting tubes, although they have the same band gap. Distinctions between, let us say ${\cal N}$$-$1 and ${\cal N}$+1 semiconducting carbon tubes (and their parent graphene ribbons) are seen in the electron states, which are obtained in Appendix B, and also, in the respective Green's functions, where these distinctions are especially prominent \cite{Lyuba4}.

Again, Eq. (\ref{7}) has to be expanded near zero-energy points in powers of $k_x$ and $\frac{2\pi(j-j^*)}{\cal N}$. In a metallic zCT, $j^*=\frac{\cal N}{3}$ and $j^*=\frac{2\cal N}{3}$ for the cross and circle points, respectively. In a semiconducting zCT, the choice of $j^*$ is explained above. As a result of the expansion, we arrive at
\begin{equation} \label{14}
\begin{array}{ll}
E^\sigma_\nu(k_x)= \pm \frac{\sqrt{3}}{2} 
\sqrt{m^{\sigma\,2}_\nu + k_x^2 },&\\
 m_\nu^\sigma = \frac{2\pi|\nu|}{{\cal N}} \left(1+ \frac{\pi \nu}{2\sqrt{3}\cal N}\right)<<1,& \sigma = 
{\rm zCTm},\\
m_\nu^\sigma =\frac{2\pi}{\cal N}\left|\nu-\frac{1}{3}\right|<<1, & \sigma = {\rm zCTs},\\
\nu=0,\pm1,...\,,&
\end{array}
\end{equation}
where extensions in labeling indicate metallic (zCTm) and semiconducting (zCTs) zigzag carbon tubes. 

Distinct from the aCT spectrum is that the $\nu$=0 band in zigzag carbon tubes is twofold degenerate, as all other $\pi$ bands near the Fermi energy. Another distinction is that within one valley associated with either cross or circle zero-energy point, it is only {\it one} value of $k_x$ that matches any energy value within the given band. In the aCT spectrum, there are {\it two} values of $k_y$ which correspond to the same energy within the $\nu$th band. 

It is worthwhile noting that the band energies in the zGR spectrum contain a linear in $\nu/{\cal N}$ term, whereas in the aCT spectrum, the correction term is quadratic in $\nu/N$, see Eq. (\ref{13}). Hence, the band spacing in aCT spectrum is nearly constant, whereas in zCT spectrum, it is essentially dependent on the band index.

From the comparison of calculations according to Eqs. (\ref{7})  and (\ref{14}), as illustrated in Fig. 5, it is apparent that retaining the $\nu$ term under parentheses in the definition of $m_\nu^{\rm zCTm}$ gives a much better agreement with the exact calculations than the use of $m_\nu^{\rm zCTm}=\frac{2\pi|\nu|}{{\cal N}}$.
The extra $\nu$-dependent term in the band energy $E^{^{\rm zCT}}_{j=j^*}(0)$ and the shift of band bottoms/tops with the increase of band index towards larger wave vectors in aCT are missed in the {\bf\it k\,$\cdot$p} approximation which is equivalent to the use of (2+1)D quantum electrodynamics formalism \cite{Ando}. As already mentioned, this difference becomes insignificant in the limit of large tube diameter.

An important conclusion that follows from Eqs. (\ref{12}), (\ref{13}), and (\ref{14}) is that near the Fermi energy, carbon tubes  can have three types of spectra: (i) with (to a good approximation) equally spaced bands, as for aCT, (ii) with regularly irregular band spacing, as for zCTm, and (iii) with an alternating band spacing, approximately, between values of $\frac{2\pi}{3\cal N}$ and $\frac{4\pi}{3\cal N}$, as for zCTs. 

Subsequent sections reveal a very close similarity between the band spectra of armchair and zigzag carbon tubes and their parent graphene ribbons, zGR and aGR, respectively. Also, one prominent exception from otherwise identical regularities will be described in detail.

\subsection{Spectrum of Armchair Graphene Ribbon}
The spectra of graphene ribbons have been persistently studied numerically at different levels, from tight-binding to ab initio \cite{GR-Fam,Kobayashi,Fujita,Nakada,W,Louie,Mint2,White}. Few attempts to attack the problem analytically \cite{White,Brey,Ch,prb2007} have been paralleled by extensive simulations of ribbon spectra. Two forthcoming sections provide a fully analytical description of aGR and zGR spectra near the Fermi energy. This description details and extends the results of works \cite{Lyuba1,Lyuba2,Lyuba3}.

It is easy to show that with minor changes, which are connected with different boundary conditions in the transverse direction, aGR spectrum repeats zCT spectrum (\ref{14}). Specifically, the condition of metallicity for aGR requires $j^*=2({\cal N}+1)/3$ to be an integer. Then, index $\nu=j-j^*=0$ corresponds to the zero-energy band. If $2({\cal N}+1)/3$ is not an integer, aGR spectrum has a gap, and the band closest to zero is either $j^*=(2{\cal N}+1)/3$ or $j^*=(2{\cal N}+3)/3$ depending on which of these two numbers is an integer. There is only one zero-energy point in the spectrum of metallic aGR, it is the cross point.

For $k_x,\nu/{\cal N}<<1$, the exact spectrum (\ref{6}) simplifies to 
\begin{equation} \label{15}
\begin{array}{ll}
E^\sigma_\nu(k_x)= \pm \frac{\sqrt{3}}{2} 
\sqrt{m^{\sigma\,2}_\nu + k_x^2 },&\\
 m_\nu^\sigma = \frac{\pi|\nu|}{{\cal N}+1} \left(1+ 
\frac{\pi \nu}{4\sqrt{3}({\cal N}+1)}\right)<<1,& \sigma = 
{\rm aGRm},\\
m_\nu^\sigma =\frac{\pi}{{\cal N}+1}\left|\nu-\frac{1}{3}\right|<<1, & \sigma = {\rm aGRs},\\
\nu=0,\pm1,...\,,&
\end{array}
\end{equation}
which repeats the spectra of metallic and semiconducting zigzag tubes, where ${\cal N}\Rightarrow 2({\cal N}+1)$. 
Another way of derivation of aGR band structure was presented in Ref. \cite{Ch}. It was concluded that "... there is no general rule of the subband index ...". The above equation constructively opposes this statement. 

The difference between the boundary conditions for aGR and zCT results in about two-times smaller band spacing in the aGR spectrum, than it was found for the zCT spectrum. Continuing this comparison we note that all $\nu$ bands in aGR spectrum are nondegenerate. This conclusion just repeats the known result \cite{Nakada}. The {\bf\it k\,$\cdot$p} approximation, where electron states in K and K$'$ valleys have to be admixed, prescribes the twofold band degeneracy in armchair graphene ribbons \cite{Brey}.  

\subsection{Spectrum of Zigzag Graphene Ribbon}

Of four basic graphene-based wires, achiral graphene ribbons and carbon tubes, only in zigzag ribbons it occurs that longitudinal and transverse motions are not separable. Analytically, this interrelation was expressed for the first time in Ref. \cite{Brey}. It has the form of a transcendent equation
\begin{equation}\label{16}
q=k_x\cot(\sqrt{3}k_xN),
\end{equation}
where $q$ and $k_x$ are, respectively, the longitudinal and transverse wave vector components referring (according to the authors of the quoted paper) to one of K points. Equation (\ref{16}) was obtained by exploiting the Dirac equation for massless fermions with the dispersion  
\begin{equation}\label{16a}
E(k_x,q)=\pm\frac{\sqrt{3}}{2} \sqrt{k_x^2+q^2}.
\end{equation}
As a particular case, Eq. (\ref{16}) appears  in a recent analysis of graphene structures with more complex edges than just zigzag or armchair \cite{Been}. However, the results of these studies are restricted by limitations of the long-wave approximation. Here, the problem is addressed on the basis of exact analogues of Eqs. (\ref{16}) and (\ref{16a}), represented in the next two equations.

The part of zGR spectrum that includes the Fermi energy is described by the minus branch of Eq. (\ref{10}), 
\begin{equation}\label{17}
\begin{array}{l}
E^{^{\rm zGR}}(k_x,k_y)= \pm 
\sqrt{1-4\cos \frac{k_y}{2}\cos\frac{\sqrt{3}k_x}{2}+ 4\cos^2\frac{k_y}{2} },
\end{array}
\end{equation}
where for each value of $0\le k_y\le\pi$, allowed values of $k_x$ has to be found from
\begin{equation}\label{18}
\frac{\sin \sqrt{3}k_xN} {\sin \sqrt{3}k_x (N+1/2)}   
= 2\cos\frac{k_y}{2}. 
\end{equation}
Note that Eq.~(\ref{16a}) is just an expansion of radicand (\ref{17}) near the cross point in powers of $k_x$ and $q = k_y-\frac{2\pi}{3}$ up to $k_x^2$ and $q^2$. Approximation (\ref{16}) follows from the exact relation (\ref{18}) after the following replacements $2\cos(\pi/3-q/2)\Rightarrow 1-\sqrt{3}q/2$ and 
$$
\begin{array}{l}
\displaystyle{
\frac{\sin \sqrt{3}k_xN} {\sin \sqrt{3}k_x (N+1/2)} }
\Rightarrow 1-\frac{\sqrt{3}k_x}{2} \cot\left(\sqrt{3}k_xN\right), 
\end{array}
$$
this cannot be justified in any rigorous way.

Equation (\ref{18}) has $N$ solutions $ k_x^{(\nu)} $, $\nu=0,1,...,N-$1, for each value of $k_y$. One of these solutions is imaginary, $\sqrt{3}k_x^{(0)} =i\delta$, if $k_y$ falls into the interval $ \frac{2\pi}{3} + q^c <ak_y\le\pi$, as illustrated in Fig. 6. Shown in this figure are graphical solutions of Eq.~(\ref{18}) for two cases, $q < q^c$ and $q>q^c$. The critical value of $q=q^c$ corresponds to $k_x^{(0)}=0$, that is 
\begin{equation}\label{19}
q^c=2\arccos\left(\frac{N}{2N+1}\right)-\frac{2\pi}{3}. 
\end{equation}
In the state with $k_x=0$ and $q=q^c$, the electron energy equals 
\begin{equation}\label{20}
\left|E^{^{\rm zGR}}\left(0,\frac{2\pi}{3}+q^c\right)\right|\equiv E^c = 1-2\cos\left(\frac{\pi}{3}+\frac{q^c}{2}\right). 
\end{equation}

\begin{figure}
\centering
\includegraphics*[width=0.8\linewidth,height=0.62\linewidth]{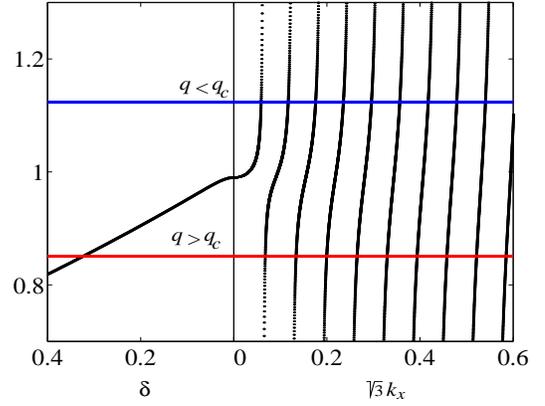}
\caption{Graphical solutions of Eq. (\ref{18}) which correspond to intersections of horizontal lines, representing the right hand side, with curves $y_1(k_x)=\frac{\sin \sqrt{3}k_xN} {\sin \sqrt{3}k_x (N+1/2)}$ and 
$y_2(\delta)=\frac{\sinh\delta N}{\sinh\delta(N+1/2)}$, $N=50$. Blue line intersects only $y_1$; red line intersects $y_1$ and $y_2$.}
\label{Fig7}
\end{figure}

For energies $|E|<E^c$, implying that $q>q^c$, the spectrum is described by Eq.~(\ref{17}), where $\sqrt{3}k_x=i\delta$, and by equation
\begin{equation}\label{21}
\frac{\sinh \delta N} {\sinh \delta (N+1/2)}   
= \cos\frac{q}{2} -\sqrt{3}\sin\frac{q}{2},\quad q>q^c.
\end{equation}
In this representation too, the quantum numbers $q$ and $\delta$ are interdependent.
The rest of the spectrum is determined by Eqs.~(\ref{17}) and (\ref{18}). 

An equivalent and in certain respects more convenient expression of the spectrum can be obtained by combining these two equations in order to exclude $k_y$ from the first of them, see Appendix C. Then, the energy of electron states becomes a function of one variable $k_x$, 
\begin{equation}\label{22}
E^{^{\rm zGR}}(k_x)= \pm \left |
\frac{\sin (\sqrt{3}k_x /2)}{\sin \sqrt{3}k_x (N+1/2)}\right |,
\end{equation}
where $k_x$ is dependent on $k_y$ via Eq.~(\ref{18}). 

Up to this point, our consideration of zGR spectrum was exact. Now, we proceed with useful approximate expressions which make apparent the electronic structure of zigzag ribbons and expose its similarity and dissimilarity with the band spectrum of armchair tubes. In this discussion, the ribbon width will be assumed large, $ N>>1$, so that $q^c= \frac{1}{\sqrt{3}(N+1/2)}$ and $E^c= (2N+1)^{-1}$. The approximate formulas to be presented give rather accurate estimates already for $N\sim 10$.

The dispersion for edge states (ES), for which $q>q^c$ and $|E|<E^c$, is described exactly but implicitly by
\begin{equation}\label{23}
E^{^{\rm ES}} (k_x{\rm=}i\delta/\sqrt{3})= \pm  
\frac{\sinh (\delta/2)}{\sinh \delta(N+1/2)}.
\end{equation}
One can see that for $N\delta>1$, the edge-state energy goes to zero exponentially with the increase of $\delta$,
\begin{equation}\label{24}
E^{^{\rm ES}}(\delta) = \pm 
2\sinh (\delta/2) e^{-\delta(N+1/2)} 
\xrightarrow[\delta >> 1]{}
\pm e^{-\delta N}. 
\end{equation}
The behavior of $E^{^{\rm ES}}(\delta(q))$, that is the edge-state dispersion shown in Fig.~7, is more complicated. Its explicit expression can be found from Eq.~(\ref{21}) under certain restrictions on $q$. 

\begin{figure}[htbp]
\includegraphics[width=0.45\textwidth]{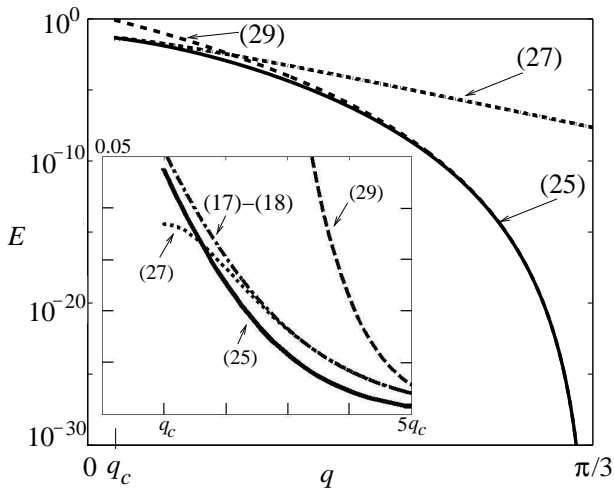}
\caption{Edge state dispersion in a $N$=10 wide zigzag ribbon. Solid curves represent exact calculations according Eqs. (\ref{18}) and (\ref{23}). Various approximations are marked by equation numbers. Curve (\ref{16}), (\ref{16a}) and curve (\ref{25}) are undistinguishable in semi-log scale. Inset: same data for small values of $q$ in non-logarithmic scale.  }
\label{Fig8}
\end{figure}

For $1\le q/q^c\lesssim2$, this dependence is reasonably described by $\delta \approx q\left[1-2 \exp(-2\sqrt{3}qN)\right]$, so that
\begin{equation}\label{25}
E^{^{\rm ES}}(q) \approx \pm {\sqrt{3}}q\exp(-\sqrt{3}qN).
\end{equation}
This approximation is close to exact calculations up to $q \approx\pi/12$. For larger wave vectors, $\pi/6<q\le\pi/3$ (or $N\delta >2$), the solution to Eq.~(\ref{21}) can be represented as
\begin{equation}\label{26}
\delta =-2\ln\left(2\sin\frac{\pi/3-q}{2}\right), 
\end{equation}
showing that with the approaching of the wave vector to its maximal value, $q\rightarrow \pi/3$ ($k_y\rightarrow\pi$), the edge state energy goes to zero as
\begin{equation}\label{27}
E^{^{\rm ES}}(q)  = \pm e^{-2N\left|\ln\left(\frac{\pi}{3}-q\right)\right|}.
\end{equation}

Thus, dispersion of edge states, which in numerical studies appears as a dispersionless band \cite{Fujita,Nakada}, is governed by exponential dependencies defined in Eqs.~(\ref{25}) (for $q^c<q<\pi/12$)  and (\ref{26}) (for $\pi/6<q<\pi/3$). The crossover between the two can approximately be designated to a region, where $\pi/12<q<\pi/6$. Analytical expressions of edge state dispersion, which are defined in the above equations, are compared with the exact results in Fig. 7. As seen, the long-wave approximation followed in Refs. \cite{Brey,Been} and illustrated by curves (\ref{16})--(\ref{16a}) and (\ref{25}) does not reproduce the dispersion of edge states in the larger part of ES band. 

Out of the ES band, $|E|>E^c$, and for $k_x,|q|<<1$, Eq.~(\ref{17}) can be rewritten as (see Appendix C \cite{Lyuba6})
\begin{equation}\label{28}
\begin{array}{ll}
{\displaystyle
E^\sigma_\nu(q) =\pm\frac{\sqrt{3}}{2}}\sqrt{m_\nu^{\sigma\,2}\left(1-\frac{\sqrt{3}}{2}q\right) +q^2},&\\
\nu=0,1,...<<N,&\sigma = {\rm zGR},\\ 
\displaystyle
m^\sigma_\nu=\frac{\pi(\nu+1/2)}{\sqrt{3}N}, &
\end{array}
\end{equation}
where $q<q^c$ within the $\nu=0$ band. Without the linear in $q$ term, the spectrum (\ref{28}) can be obtained from Eqs. (\ref{16}) and (\ref{16a}). 

As shown by many numerical calculations (but never proved analytically), with an exception of the ES band energy interval, zGR and aCT have a very similar band spectra. Both spectra have nearly equally spaced bands with the bottoms/tops which are pronouncedly shifted towards larger wave vectors with the increase of band index. The band spacing in zGR spectrum is two times smaller, $\frac{\pi}{2N}$, than it is in the aCT spectrum. Therefore, aCT bands are in-between pairs of zGR bands. The band structure parameters for zGR, band energies and positions of bands are given by the same equations as for aCT, Eq.~(\ref{12a}) and Eq.~(\ref{13}), where $m_\nu^{^{\rm aCT}}$ should be replaced by $m_\nu^{^{\rm zGR}}$. All these features can be seen in Fig. 5 that compares the full aCT and zGR spectra of $\pi$ electrons, as well as their details near zero-energy points. 

So, for $|E|>E^c$, the zGR spectrum can be obtained from the aCT spectrum by the "mass" scaling, and the other way round. In the narrow energy interval $2E^c$ near the Fermi energy, aCT and zGR spectra are qualitatively different. This is connected with the appearance of ES band in zigzag graphene ribbons. As shown, the dispersion within this band is exponential and goes to zero, when the wave vector approaches its maximal value. The {\bf\it k\,$\cdot$p} approximation, as it was used in Ref. \cite{Brey}, leads to a "reversed" picture, where ES energy tends to zero when the wave vector goes to zero. 

To wind up this section, we note that only marginal changes are required to apply the above scheme of calculations to carbon tubes and graphene ribbons of finite length, taking into account their termini. In particular,
the gap between HOMO (the highest occupied molecular orbital) and LUMO (the lowest unoccupied molecular orbital)
is equal to \cite{MO}
\begin{equation}\label{29}
E^\sigma_{_{\rm H{\rm-}L}} =2 \left \{
\begin{array}{lll}
0, & \frac{{\cal N}+1}{3} \;{\rm integer},  & \\
&&\sigma ={\rm aCTS}, \\
\frac{\pi}{2\sqrt{3}({\cal N}+1)},&{\rm otherwise}, &\\
&&\\
0& {\rm even}\;{\cal N},& \\
&&\sigma ={\rm zCTS},  \\
e^{-2N\left|\ln\frac{\pi}{\cal N}\right|} & {\rm odd}\;{\cal N},&\\
&&\\
e^{-2N\left|\ln\frac{\pi}{{\cal N}+1}\right|}&&\sigma ={\rm GS},
\end{array} \right. 
\end{equation}
where $\cal N$ and $N$ is the length of aCT and zCT segments denoted as aCTS and zCTS, respectively; GS abbreviates armchair ($N>>{\cal N}>>1$) and zigzag (${\cal N}>>N>>1$) graphene strips. HOMO-LUMO gap values and other characteristics which have been discussed here, are summarized in Table 1.

\subsection{Relativistic Analogy}

The band structure of carbon tubes and graphene ribbons can be summarized in a single line \cite{Note2},
\begin{equation}\label{30}
E^\sigma_\nu(k_\nu^\sigma) = \pm
\sqrt{m^{\sigma\,2}_\nu + k_\nu^{\sigma\,2}}, 
\end{equation}
where $k^\sigma_\nu$ has the meaning of dimensionless wave vector, $m^\sigma_\nu$ (only here, in units of $\frac{\sqrt{3}}{2}|t|$) is the $\nu$th band energy, and index $\sigma$ specifies the structure and the $\nu$th band minimum/maximum in the 1D $k$ space $\bar k^\sigma_\nu$ as follows 
\begin{equation}\label{31}
m^\sigma_\nu =
\begin{array}{l}
\frac{\pi|\nu |}{{\cal N}}\left(1{\rm+} \frac{\pi \nu }{4\sqrt{3}\cal N}\right) , \\
\frac{\pi|\nu -1/3|}{{\cal N}} ,\\
\frac{\pi(\nu +1/2)}{\sqrt{3}N},\\ 
\frac{2\pi|\nu|}{{\cal N}}\left(1{\rm+} \frac{\pi \nu }{2\sqrt{3}\cal N}\right), \\
\frac{2\pi|\nu -1/3|}{{\cal N}} ,\\ 
\frac{2\pi|\nu|}{\sqrt{3}N},  
\end{array}
\quad
\bar k^\sigma_\nu =
\begin{array}{l}
0, \\
0,\\ 
\frac{2\pi}{3} {\rm+} \frac{\sqrt{3}}{4} \left(m^{\rm zGR}_\nu\right)^2, \\\\ 
0,\\
0,\\
\frac{2\pi}{3}{\rm+} \frac{\sqrt{3}}{4} \left(m^{\rm aCT}_\nu\right)^2,
\end{array} 
\end{equation}
$\sigma$ = aGRm, aGRs, zGR, zCTm, zCTs, aCT, from top to bottom.  Equation (\ref{30}) is valid for $k^\sigma_\nu,m_\nu^\sigma<<1$; $\nu=0,\pm1,...$ for all structures, except zGR in which case $\nu=0,1,...$ .

One can see that within each band (except the ES band) the dispersion relation (\ref{30}) has the form of 1D relativistic energy-momentum relation in its conventional representation with the speed of light equal to unity \cite{AM}. Thus, from the point of view of energy and momentum conservation laws, quasiparticles in the $\nu$=0 conduction/valence bands of metallic carbon tubes and graphene ribbons (except zGR) should behave as 1D massless Dirac fermions (or neutrinos/antineutrinos), whereas in $\nu$$\ne$0 bands they should behave as relativistic particles having mass $m_\nu^\sigma$. 

Perfect penetration of electrons/holes with linear dispersion into a classically impenetrable region has been noticed in several publications \cite{Ando1,Fal,Kats1}. 
More recently \cite{Lyuba5}, it was shown that by passing from the exact tight-binding description to approximation (\ref{30}), expressions for the probabilities of tunneling through a potential step and through a potential barrier within the $\nu$th band (and inter-band scattering prohibited) exactly coincide with formulas, derived almost eighty years ago for tunneling of relativistic particles \cite{Klein,Klein1}. It is really amazing that by predicting the penetration probability for massive relativistic particles, Oskar Klein predicted the probability of transmission of charge carriers through n/p junctions in alternant macromolecules and, in particular, in graphene, a material unknown in his time. 

Equations (\ref{30})  and (\ref{31}) make obvious that spectra of achiral carbon tubes and graphene ribbons can be divided into three groups: (i) metallic spectra with equally spaced bands, case of aCT and zGR; (ii) metallic spectra with an irregular band spacing, where in case of zCT (aGR)  it can be any fraction of $\frac{2\pi}{\cal N}$ ($\frac{\pi}{{\cal N}{\rm+1}}$) and also, larger than that; and (iii) semiconducting spectra with a band spacing, alternating between $\frac{2\pi}{3\cal N}$ and $\frac{4\pi}{3\cal N}$, and between $\frac{\pi}{3\cal N}$ and $\frac{2\pi}{3{\cal N}{\rm+1}}$ in cases of semiconducting zigzag tubes and armchair ribbons, respectively. 
These differences in the band structures along all {\it cis} and all {\it trans} carbon chains reflect in a quantitative manner the nonequivalence of armchair and zigzag directions in the honeycomb lattice structures. In the infinite graphene sheet these differences disappear.

\section{Quantum Conductance}\label{Sec5}

Within the framework of the Landauer approach \cite{Land,But1,But2,Note3}, the zero-temperature ohmic conductance of an ideal wire is equal to
\begin{equation}\label{32}
G^\sigma(E)= G_0\sum_{\nu}g^\sigma_\nu T^\sigma_{\nu}(E),
\end{equation}
where $G_0=2e^2/h$ is conductance quant, $g^\sigma_\nu$ is the band degeneracy (spin degeneracy 2 is included into $G_0$), and transmission coefficient $T^\sigma_\nu$ is zero or unity, depending on whether the $\nu$th band is open or closed for charge carriers with energy $E$. 

With an account to Eq. (\ref{30}), $T^\sigma_\nu(E) = \Theta(E-\frac{\sqrt{3}}{2} m^\sigma_\nu)$ for conduction bands, and $T^\sigma_\nu(E) = \Theta(|E- \frac{\sqrt{3}}{2} m^\sigma_\nu|)$ for valence bands; $\Theta(x)$ is the Heaviside step function. The values of $g^{\rm aCT}_\nu$ and $g^{\rm zGR}_{\nu\ne0}$ represented in Table 1 must be doubled because electron/hole states with $\pm k_\nu^{\rm aCT}\ne0$ and $\pm k_{\nu\ne0}^{\rm zGR}\ne0$ are degenerate. This "rule" was earlier noticed in the numerical study of GR conductance \cite{Peres1}. 

The electron/hole conductance of armchair and zigzag carbon tubes and their parent graphene ribbons has thus the form of a ladder, symmetrically ascending with the increase of energy for electrons, and with the decrease of energy for holes. For the charge carrier energy that falls in-between the $\nu$th and ($\nu$+1)the bands, the wire conductance equals 
\begin{equation} \label{33}
G^\sigma(E)= G_0\left\{
\begin{array}{ll}
\nu& {\rm aGRm,\,aGRs},\\
2\nu+1& {\rm zGR},\\
2\nu& {\rm zCTm,\,zCTs},\\
2(2\nu+1)& {\rm aCT},
\end{array}
\right. 
\end{equation}
The width of steps repeats the band spacing $m_\nu^\sigma$ defined in Eq. (\ref{31}) in units of $\frac{\sqrt{3}}{2}|t|$. The height of the $\nu$th ladder step is $G_0$ times band degeneracy as explained above. 

In the aforementioned study \cite{Peres1}, the step width was calculated numerically. For $G^{^{\rm zGR}}(E)$, the same expression was suggested, whereas  $G^{^{\rm aGRm}}(E)$ according to the quoted paper is equal to $2\nu G_0$; for $G^{^{\rm aGRs}}(E)$ no expression  was presented. The possible reason for the extra factor of two in the conductance of metallic armchair ribbons is explained bellow.  

As long as dispersion (\ref{30}) is valid, the conductance ladders of basic graphene wires can be classified in the following three types: (i) with {\it regular} step width for always metallic aCT and zGR wires; (ii) with {\it irregular} step width for metallic zCT and aGR; and (iii) with {\it alternating} step width for semiconducting zCT and aGR. These characteristics distinguish graphene ribbons and carbon tubes as 1D quantum wires from the conductance ladders known for 2DEG channels \cite{Been3}.

The appearance of $G^\sigma(E)$ depends on the energy scale which, in its turn, is determined by the ribbon width (tube circumference). Experimentally, this can result in different observations, from ladders, which might be fully or partly resolved, to unresolved ladders. In the latter case, that is when the discrete behavior of quantum conductance cannot be resolved, all ladders are smoothed out into straight lines with the same slope and zero-energy values equal to $G^{^{\rm GR}}(0)=G_0$ and $G^{^{\rm CT}}(0)=2G_0$ for graphene ribbons and carbon tubes, respectively. 

Returning to the aGRm conductance, we remind that in this case, the step width equals $\frac{\pi|\nu |}{{\cal N}}\left(1{\rm+} \frac{\pi \nu }{4\sqrt{3}\cal N}\right)$, see Eq. (\ref{31}). Therefore, for large $N$, quite many pairs of bands near the Fermi energy have very close energies, i.e., they are apparently degenerate. Then, $G^{^{\rm aGRm}}(E)= 2\nu$, if $\nu<<N$. Obviously, under the same conditions $G^{^{\rm zCTm}}(E)= 4\nu$. 

\section{Synopsis}\label{Sec6}

A new methodology of analytical modeling of $\pi$ electron spectrum of graphene and its daughter lattices, achiral carbon tubes and graphene ribbons, has been presented. It is based on the tight-binding model of graphene as a macromolecule with armchair- and zigzag-shaped boundaries. The exact solution of the Schr{\"o}dinger problem, the spectrum and wave functions, have been obtained and illustrated by a number of examples. Several spectral features, which were previously accessible only for numerical calculations, have received an adequate analytical description in terms of elementary functions. In comparison with the understanding based on the 2D graphite band structure, the macromolecule model is shown to be more relevant and more beneficial. It provides a consistent description of graphene electronic properties. Presenting the full details of this model sheds light on the intimate interrelation between graphene, acenes, and other conjugated oligomers. In general, this model gives more comprehensive picture of what may be called relativistic appearance of graphene.\\\\

\begin{acknowledgments}
The author is deeply thankful to Bo Liedberg for providing excellent working conditions in his group, Lyuba Malysheva for
 critical reading this manuscript and for numerous discussions and derivations which contributed significantly to
 the presented study of graphene, and to Linda Wylie for help in editing the manuscript. Financial support of the
 work from Swedish AEA and TSN is gratefully acknowledged.
\end{acknowledgments}

\appendix

\section{Spectrum}
By neglecting the overlap between $\pi$ orbitals at neighboring C atoms, $\langle m,n,\alpha |m',n',\alpha'\rangle=\delta_{mm'}\delta_{nn'}\delta_{\alpha\alpha'}$, Eqs. (\ref{1}) and (\ref{2}) can easily be transformed into a set of four equations 
\begin{equation}\label{A1}
E\psi_{m,n,l}=\psi_{m,n-1,r}+\psi_{m,n,\lambda}+\psi_{m+1,n,\lambda},
\end{equation}
\begin{equation}\label{A2}
E\psi_{m,n,\lambda}=\psi_{m,n,l}+\psi_{m-1,n,l}+\psi_{m,n,\rho},
\end{equation}
\begin{equation}\label{A3}
E\psi_{m,n,\rho}=\psi_{m,n,r}+\psi_{m-1,n,r}+\psi_{m,n,\lambda},
\end{equation}
\begin{equation}\label{A4}
E\psi_{m,n,r}=\psi_{m,n+1,l}+\psi_{m,n,\rho}+\psi_{m+1,n,\rho}.
\end{equation}
For the $N$$\times$$\cal N$ sheet of graphene, these equations are to be solved with open boundary conditions which read
\begin{equation}\label{A4a}
\begin{array}{l}
\psi_{0,n,l}=\psi_{0,n,r}=\psi_{{\cal N}+1,n,l}=\psi_{{\cal N}+1,n,r}=0,\\
\psi_{m,0,r}=\psi_{m,N+1,l}=0.
\end{array}
\end{equation}
 
Coefficients $\psi_{m,n,\lambda}$ and $\psi_{m,n,\rho}$ can be expressed in terms
of $\psi_{m,n,l}$ and $\psi_{m,n,r}$ as
\begin{equation}\label{A5}
(E^2{\rm -}1)\psi_{m,n,\lambda}{\rm =}E\left (\psi_{m,n,l}{\rm +}\psi_{m{\rm -}1,n,l}\right )+\psi_{m,n,r}+\psi_{m{\rm -}1,n,r},
\end{equation}
\begin{equation}\label{A6}
(E^2{\rm -}1)\psi_{m,n,\rho}{\rm =}E\left (\psi_{m,n,r}{\rm +}\psi_{m{\rm -}1,n,r}\right )+ \psi_{m,n,l} +\psi_{m{\rm -}1,n,l}.
\end{equation}
Exploiting these two equations in (\ref{A1}) and (\ref{A4}), we obtain a reduced set, involving only $\psi_{m,n,l}$ and $\psi_{m,n,r}$. This set has the form
$$
E(E^2{\rm -}1)\psi_{m,n,l}{\rm =}(E^2{\rm -}1)\psi_{m,n{\rm -}1,r}{\rm +}E\left ( 2\psi_{m,n,l}{\rm +}\psi_{m{\rm -}1,n,l} \right. 
$$
\begin{equation}\label{A7}
{\rm +} \left. \psi_{m{\rm +}1,n,l} \right ) {\rm +}2\psi_{m,n,r}{\rm +}\psi_{m{\rm -}1,n,r}
 {\rm +}\psi_{m{\rm +}1,n,r},
\end{equation}
$$ 
E(E^2{\rm -}1)\psi_{m,n,r}{\rm =}(E^2{\rm -}1)\psi_{m,n{\rm +}1,l}{\rm +} E\left (2\psi_{m,n,r}{\rm +}\psi_{m{\rm -}1,n,r} \right . 
$$
\begin{equation}\label{A8}
{\rm +} \left. \psi_{m{\rm +}1,n,r} \right ){\rm +}2\psi_{m,n,l}{\rm +}\psi_{m{\rm -}1,n,l} +\psi_{m{\rm +}1,n,l}.
\end{equation}

Now, it is convenient to represent wave function coefficients in the form (\ref{3}). By substituting it into Eqs. (\ref{A7}) and (\ref{A8}) and performing standard algebra, we arrive at 
\begin{equation}\label{A9}
\begin{array}{ll}
\phi_{n,\alpha}^j  =g_{\alpha,l}^j\phi_{n-1,r}^j+ g_{\alpha,r}^j \phi_{(n+1),l}^j,&\alpha=l,r,\\
\phi_{0,r}^j  = \phi_{N+1,l}^j  =0,
\end{array}
\end{equation}
where $g_{l,r}^j =g_{r,l}^j=4\cos^2(\xi_j/2) 
{\cal D}_j^{-1}$, $\xi_j=\frac{\pi j}{{\cal N}+1}$,
 $g_{l,l}^j$ = $g_{r,r}^j$ = $E[E^2-1-4\cos^2(\xi_j/2)] {\cal
D}_j^{-1}$, and
\begin{equation}\label{A10}
{\cal D}_j= [E^2-E-4\cos^2(\xi_j/2)][E^2+E-4\cos^2(\xi_j/2)].
\end{equation}
The latter equation is nothing else but the determinant of the spectral problem for linear acenes \cite{Lyuba4}. Under replacement $\xi_j\Rightarrow k_y$, Eq. (\ref{A10}) converts into the dispersion relation for polyacene \cite{Kiv}.

Formally the same equation as Eq. (\ref{A9}) appears in the theory of M-oligomers \cite{jcp107,Chapt} which are linear molecules consisting of $N$ monomers M coupled to each other via left and right binding atoms, as illustrated in Fig. 8. Monomer M can be described by the Green's function $G^{\rm M}_{\alpha,\alpha'}=\langle\alpha|(I-H^{\rm M})^{-1}|\alpha'\rangle$, where $H^{\rm M}$ is one-particle Hamiltonian in the tight-binding representation. Otherwise, M is an arbitrary complex of $N_{\rm M}$ atoms. 

In Eq. (\ref{A9}), $g_{\alpha,\alpha'}^j$ associates with the Green's function matrix element of a hypothetical monomer indicated by the dashed frame in Fig. 8. Thus, all relations which follow from Eq. (\ref{A9}), can be exploited in the present context. In general terms, the description of M-oligomers with the use of Eq. (\ref{A9}) represents a generalization of the Lennard-Jones theory of polyenes, (M)$_N$, $N_{\rm M}$ = 2, M = C=C \cite{LJ}. 

\begin{figure}
\includegraphics[width=0.47\textwidth,height=0.57\linewidth]{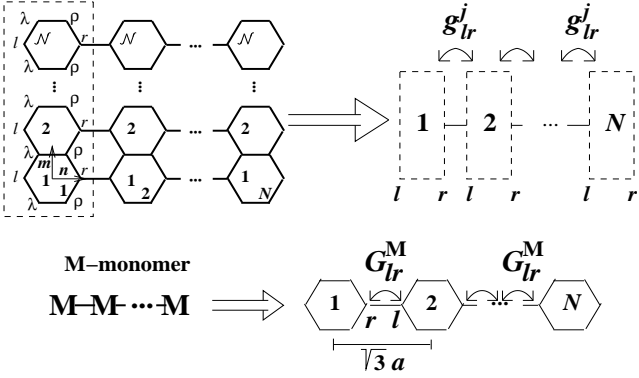}
\caption{$N$$\times$$\cal N$ graphene sheet in Fig. 1, represented as a sequence of dashed-frame boxes, illustrates the analogy with M-oligomer structure exemplified by $N$-long oligomer of polyparaphenylene. In Eq. (\ref{A9}), $g_{l,r}^j$ plays the role of monomer Green's function matrix element referring to monomer binding atoms $l$ and $r$.}
\label{Fig11}
\end{figure}

The $\pi$ electron spectrum of conjugated M-oligomers is determined by two equations \cite{jcp107,Chapt}. One of them relates the state energy $E$ with the wave vector expressed in units of periodicity of the oligomer chain, 
\begin{equation}\label{A11}
\cos\kappa = f(G^{\rm M}),
\end{equation}
where $f(G^{\rm M})$ is a functional of monomer Green's function matrix elements,
\begin{equation}\label{A12}
f( G^{\rm M} )\equiv \frac{1}{2G^{\rm M}_{l,r}}\left(1+ G^{\rm M\,2}_{l,r}
-G^{\rm M\,2}_{l,l}\right).
\end{equation}
In our case, $\kappa=\sqrt{3}k_x$, $k_x$ is in units of $a^{-1}$. The other equation determines allowed values of the wave vector which must satisfy equation
\begin{equation}\label{A13}
\frac{\sin \kappa N}{\sin \kappa(N+1)}=-\frac{G^{\rm M}_{l,r}}
{G^{\rm M\,2}_{l,l} - G^{\rm M\,2}_{l,r}}.
\end{equation}
Equations (\ref{4})  and (\ref{5}) follow from Eqs. (\ref{A12})--(\ref{A14}) after replacing $G^{\rm M}_{l,l}$ and $G^{\rm M}_{l,r}$ by $g^j_{l,l}$ and $g^j_{l,r}$ and some algebra. 

It is easy to see that the analogy with M-oligomers is straightforwardly applicable to segments of zigzag carbon tubes. The only difference is that in Eqs. (\ref{A9}) and (\ref{A10}), we have to replace $\xi_j=\frac{\pi j}{{\cal N}+1}$ by $\xi_j=\frac{2\pi j}{\cal N}$, $j=0,1,...,\cal N$$-$1. This is to say that the role of hypothetical monomer (framed box in Fig. 8) is played by cyclacene chain (instead of linear acene chain). 

In this case, applying Eqs. (\ref{A11})--(\ref{A13}), one obtains
\begin{equation}\label{A14}
E_j^{\pm\;2}=
1\pm4\left|\cos \frac{\pi j}{\cal N}\right|  \cos \frac{\kappa_j^\pm}{2}+
4\cos^2 \frac{\pi j}{\cal N} , 
\end{equation}
where for the sign + and $-$, and each value of $j$, except $j={\cal N}/2$, $N$ values of $\kappa_j^\pm$, $ \kappa^\pm_{j,\nu_j}$, $\nu_j=0,1,...,N$-1, are solutions to 
\begin{equation}\label{A15}
\frac{\sin \kappa_j^\pm N}{\sin \kappa_j^\pm (N+1/2)}=\mp 2\left |\cos \frac{\pi j}{\cal N}\right |
\end{equation}
for $j<[{\cal N}/2]$,  and 
\begin{equation}\label{A16}
\frac{\sin \kappa_j^\pm N}{\sin \kappa_j^\pm (N+1/2)}=\pm 2\left |\cos \frac{\pi j}{\cal N}\right |
\end{equation}
for $j>[{\cal N}/2]$. Here, $[x]$ is the nearest integer function. 

Equations ~(\ref{4}) and (\ref{A14}) were used in Sec. 2 to obtain dispersion in graphene daughter structures. The wave functions for these structures are discussed next. 

\section{Wave Functions \cite{N3}}

The wave function (\ref{2}), which corresponds to the energy $E_j^\pm(\kappa^\pm_{j,\nu_j})$, can be represented as
\begin{equation}\label{B1}
\Psi_{j,\kappa_j^\pm} = A_{j,\kappa_j^\pm} \sum_{m}\sum_{n}\sum_{\alpha=l,r,\lambda,\rho}
\psi^{j,\kappa_j^\pm}_{m,n,\alpha} |m,n,\alpha\rangle,
\end{equation}
where $A_{j,\kappa_j^\pm}$ is the normalization constant,
\begin{equation}\label{B2}
\psi^{j,\kappa_j^\pm}_{m,n,\alpha}=\left\{
\begin{array}{rl}
-s_1&\sin\frac{\pi jm}{({\cal N}+1)}\sin[\kappa^\pm_j (N+1-n)],\\ 
&\sin\frac{\pi jm}{({\cal N}+1)}\sin(\kappa^\pm_j n),\\ 
\pm &\sin\frac{\pi j(m-1/2)}{({\cal N}+1)} \sin[\kappa^\pm_j (n-1/2)] , \\
\mp  s_1 &\sin\frac{\pi j(m-1/2)}{({\cal N}+1)}\sin[\kappa^\pm_j (N+1/2-n)] , 
\end{array}\right.
\end{equation}
$s_1\equiv {\rm sign} \left[\sin \kappa_j(N+1/2) \right]$, and $\alpha =l,r,\lambda,\rho$ from up to down. For each $j$, there is $N$ solutions of Eq. (\ref{5}) with plus and minus in its right hand side.

Thus defined, coefficients (\ref{B2}) satisfy boundary conditions (\ref{A4a}). For segments of armchair carbon tubes (aCTS), we have, instead of Eq. (\ref{A4a}),
\begin{equation}\label{B3}
\begin{array}{l}
\psi_{0,n,l}=\psi_{0,n,r}=\psi_{{\cal N}+1,n,l}=\psi_{{\cal N}+1,n,r}=0,\\
\psi_{m,1,\alpha}=\psi_{m,N+1,\alpha}.
\end{array}
\end{equation}
With these boundary conditions, it is convenient to represent energies of $\pi$ electron states in aCTS as follows
\begin{equation}\label{B4}
E^{^{\rm aCTS}}_{jj'}=|z_{jj'}|, 
\end{equation}
\begin{equation}\label{B5}
z_{jj'}\equiv
\left\{
\begin{array}{l}
2\cos\frac{\pi j}{2({\cal N}+1)}\pm s_2e^{-i\pi \frac{j'}{N}},\\
j=1,2,...,{\cal N},\, j' = 0,1,...,N{\rm-1},
\end{array}\right.
\end{equation}
and $ s_2 = {\rm sign}(\cos\frac{\pi j'}{N})$. 

In these notations, eigen states of aCTS take the form
\begin{equation}\label{B6}
\Psi_{j,j'} = A_j \sum_{m}\sum_{n}\sum_{\alpha=l,r,\lambda,\rho}
\psi^{j,j'}_{m,n,\alpha} |m,n,\alpha\rangle,
\end{equation}
where
\begin{equation}\label{B7}
\psi^{jj'}_{m,n,\alpha} = 
\left\{
\begin{array}{rl}
\displaystyle \pm s_2 \frac{z_{jj'}}{|z_{jj'}|}& 
\displaystyle \sin\frac{\pi jm}{({\cal N}+1)}e^{2i\pi\frac{j'(n-1/2)}{N}},\\ 
&\displaystyle \sin\frac{\pi jm}{({\cal N}+1)} e^{2i\pi\frac{j'n}{N}} ,\\ 
\displaystyle \pm  s_2&
\displaystyle  \sin\frac{\pi j(m-1/2)}{({\cal N}+1)} e^{2i\pi \frac{j'(n-1/2)}{N}}, \\
\displaystyle \frac{z_{jj'}}{|z_{jj'}|} &
\displaystyle \sin\frac{\pi j(m-1/2)}{({\cal N}+1)}e^{2i\pi \frac{j'n}{N}},
\end{array}\right.
\end{equation}
with the same correspondence between $\alpha$ and $l$, $r$, $\lambda$, and $\rho$ as in Eq. (\ref{B2}).

The wave functions for other structures discussed in the main text can be obtained via obvious modifications of Eqs. (\ref{B1}) and (\ref{B2}) or Eqs. (\ref{B6}) and (\ref{B7}).

\section{Proof of Equation (\ref{28}) }
Equation  (\ref{18}) represents the relation between $\cos(k_y/2)$ and $k_x$ in the graphene sheet framed by two armchair-shaped and two zigzag-shaped boundaries, as shown in Fig. 1. The substitution of this specific expression of $\cos(k_y/2)$ into Eq. (\ref{17}) and usage of a chain of trigonometric identities yields
\begin{eqnarray}\label{C1}
&& 1-4 \cos\frac{k_y}{2}\cos\frac{\sqrt{3}k_x}{2}+4\cos^2\frac{k_y}{2}
\nonumber\\
&& =\frac{\sin^2 (\sqrt{3}k_x/2)}{\sin^2 \sqrt{3}k_x(N+1/2)}.
\end{eqnarray}
This gives (\ref{22}).

All $N$ roots of Eq. (\ref{18}), $0\le k_x^\nu\le \pi$, $\nu = 0, 1, \dots, N-1$, satisfy the following inequality
\begin{equation}\label{C2}
0<\frac{\sin\sqrt{3}k^\nu_x N}{\sin \sqrt{3}k^\nu_x(N+1/2)} <2.
\end{equation}
The $\nu$th root is bounded from below by $\frac{\pi \nu}{\sqrt{3}N}$. Therefore, we can redefine 
$k_x^\nu$ as follows,
\begin{equation}\label{C3}
\sqrt{3}k_x^\nu=\frac{\pi}{N}(\nu+\Delta_\nu). 
\end{equation}

It can be proved that $0<\Delta_\nu<1$. Moreover, it turns out that for sufficiently small $\nu << N$, $\Delta_\nu\approx \frac{1}{2}$. In this approximation, the substitution of (\ref{C3}) into the left hand side of Eq. (\ref{18}) leads to
\begin{equation}\label{C4}
 \frac{\sin \sqrt{3}k^\nu_x N} {\sin \sqrt{3}k^\nu_x(N+1/2)}
 =\left [1+\frac{\pi(\nu+\Delta_\nu)}{2N} \; {\rm cotan} (\pi \Delta_\nu) \right ]^{-1} .
\end{equation}
On the other hand, for $q<<1$, 
\begin{equation}\label{C5}
2\cos\frac{k_y}{2}\approx 1-\sqrt{3}q/2-q^2/8. 
\end{equation}
By equating these two expressions of the same quantity, we obtain
\begin{equation}\label{C6}
 {\rm tan} (\pi \Delta_\nu) \approx \frac{\pi(\nu+\Delta_\nu)}{\sqrt{3}qN}.
\end{equation}
The latter equation can be solved approximately, thus, solving the initial problem.

Let us assume that $|q|<\frac{\pi(\nu+1/2)}{\sqrt{3}N}$. Then,
\begin{equation}\label{C7}
k^\nu_x \approx \frac{\pi (\nu+1/2)}{\sqrt{3}N}- \frac{q}{ \pi(\nu+1/2)},
\end{equation}
showing that real solutions of Eq. (\ref{18}) depend linearly on the wave vector $q<<1$. This is in contrast with the exponential dependence of the imaginary solutions discussed in Sec. IIID. If $q\lesssim q^c$,
\begin{equation}\label{C8}
k^\nu_x \approx \frac{\pi (\nu+1/2)}{N\sqrt{3}}.
\end{equation}
Equations (\ref{C7}) and (\ref{C8}) can also be obtained from Eq.~(\ref{16}).

These results prove Eq. (\ref{28}) and justify a simple picture of zGR spectrum above and below the band of edge states (but not far from the Fermi energy), where $q$ and $k_x$ can be considered as independent quantum numbers, particularly, near the bottoms/tops of conduction/valence bands.

\end{document}